\documentclass[12pt]{article}
\usepackage{makeidx}
\usepackage{epsf}

\newcommand{\beq}{\begin{equation}}
\newcommand{\eeq}{\end{equation}}
\newcommand{\bq}{\begin{quotation}}
\newcommand{\eq}{\end{quotation}}
\newcommand{\bc}{\begin{center}}
\newcommand{\ec}{\end{center}}
\begin{document}

\title{The soliton stars evolution.}
\author{\textit{Ilona Bednarek} \\
\textit{Institute of Physics,} \\
\textit{Silesian University,}\\
\textit{Uniwersytecka 4, 40-007 Katowice, Poland} \and \textit{Ryszard
Ma\'{n}ka} \\
\textit{Institute of Physics,} \\
\textit{Silesian University,}\\
\textit{Uniwersytecka 4, 40-007 Katowice, Poland}}
\date{}
\maketitle

\begin{abstract}
The evolution of a soliton star filled with fermions is studied
in the framework of general relativity. Such a system can be described
by the surface tension $\sigma$, the bag constant $B$, and the fermion
number density $\rho_{0}$. Usually one of these parameters
prevails in the system and thus affects the spacetime inside the soliton.
Whether it is described by Friedman or de Sitter metric depends on the prevailing parameter.
The whole spacetime is devided by the surface of the soliton
into the false vacuum region inside the soliton and the true vacuum region
outside, the latter being described by the Schwarzschild line element.
The aim of this paper is to study the equations of motion of the domain wall in two cases.
In the
first case the de Sitter
metric describes the interior in the first case, and in the second case it is replaced
by the Friedman metric.
In both of them the Schwarzschild metric is outside the soliton.
From the analysis of obtained equations one can draw conclusions concerning
further evolution of a soliton star.
\end{abstract}


\newpage

\vskip13cm





In this paper we study the evolution of a soliton stellar configuration.
Let us consider the $SU_{c}$ colour QCD plus a phenomenological colourless
scalar field $\Phi$ describing the colour confinement.
The whole system is described by the scalar field $\Phi $, the fermion quark field
$\psi _f$, the gluon field $A_\mu ^a$ and the gravitational field $g_{\mu
\nu }$. The generalized QCD Lagrangian function takes the form: 
\begin{eqnarray}
\mathcal{L} &=&-\frac 14(1-\frac{\Phi ^2}{\Phi _\upsilon ^2})F_{\mu \nu
}^cF^{c\mu \nu }+\frac 12\partial _\mu \Phi \partial ^\mu \Phi -U(\Phi ) \\
&&+\overline{\psi }_f(i\gamma ^\mu D_\mu -m_f)\psi _f-g\overline{\psi }%
_f\psi _f\Phi -\frac 1{2\kappa }R  \nonumber  \label{eq:lag}
\end{eqnarray}
and consists of two parts; the basic QCD Lagrangian supplemented by the
Lagrangian of the dynamical scalar field.
In the Lagrangian function (\ref{eq:lag}) $R$ is the curvature scalar and
\begin{equation}
\kappa =8\pi G\sim \frac 1{M_{Pl}^2}
\end{equation}
$G$  the gravitational constant and $U(\Phi )$  the potential. Due to
nonlinearity of the potential $U(\Phi )$ the Lagrangian (\ref{eq:lag})
leads to
nontopological soliton solutions. The potential possesses the polynomial
form of the fourth order \cite{wi:nont} 
\begin{equation}
U(\Phi )=\frac a{2!}\Phi ^2+\frac b{3!}\Phi ^3+\frac c{4!}\Phi ^4+B.
\end{equation}
The constants $a,b,c$ are parameters which are adopted in the restriction to
enable fitting of the static properties of hadrons. The bag constant $B$ is
a measure of the pressure of the physical vacuum on the perturbative one.
The potential possesses two minima, one at $\Phi =0$ which is a local
minimum associated with the perturbative vacuum state and the second, the
absolute minimum, at 
\begin{equation}
\Phi _c=\frac{3b}{2c}[1+(1-\frac{8ac}{b^2})^{\frac 12}]
\end{equation}
corresponding to the physical vacuum. The form of the potential is presented
on Fig.1. The soliton solution exists between two limiting cases $B=0$ which
represents the degenerate vacuum state and $B=B_{max}$ when the local
minimum and maximum overlap and an inflection point appears. This is the
moment when the perturbative and physical vacuum states become metastable
and  also the limiting case in which the soliton solution still exists.\\
The action
\begin{equation}
S=\int {d^4x\sqrt{-g}\mathcal{L}}  \label{eq:act}
\end{equation}
with $\mathcal{L}$ defined as (\ref{eq:lag}) leads to the appropriate
equations of motion.
We are looking for  spherically symmetric solution with the metric
\begin{equation}
ds^2=g_{\mu \nu}dx^{\mu}dx^{\nu}=-e^{\nu}dt^2+e^{\lambda}dr^2+r^2(d\theta^2+\sin^2\theta d\varphi^2)
\end{equation}
Variation of the action (\ref{eq:act}) with respect to
metric $g_{\mu \nu }$  gives the four-dimensional Einstain equations
\begin{equation}
R_{\mu \nu }-\frac 12g_{\mu \nu }R=\kappa T_{\mu \nu }
\end{equation}
with the energy-momentum tensor $T_{\mu \nu }$ 
\begin{equation}
T_{\mu \nu }=T_{\mu \nu }(\Phi )+T_{\mu \nu }(\psi )
\end{equation}
where 
\begin{equation}
T_{\mu \nu }(\Phi )=\partial _\mu \Phi \partial _\nu \Phi -g_{\mu \nu }%
\mathcal{L}(\Phi )
\end{equation}
is the scalar field energy-momentum tensor and $T_{\mu \nu }(\psi )$  its
fermion counterpart.\\ The soliton bag model in the thin-wall approximation
describes the quark star model with a surface tension. The whole spacetime
is devided by the surface of the soliton into the false vacuum region inside
the soliton and the true vacuum region outside, the latter being described
by the Schwarzschild line element. The aim of this paper is to study the
dynamics of the domain wall in two cases. In both of them the Schwarzschild
metric is outside the soliton.\\ Subscripts $\pm $ refer to values of the
corresponding coordinates in the region outside and inside the soliton. 
\begin{equation}
ds_{+}^2=-(1-\frac{r_g}r)dt_{+}^2+\frac 1{(1-\frac{r_g}{r_{+}}%
)}dr_{+}^2+r_{+}^2(d\theta ^2+sin^2\theta d\phi ^2)
\end{equation}
where $r_g=2GM$ with $M$ being the total soliton mass.\\ The de Sitter metric
describes the interior of the soliton in the case when the value of the
cosmological constant $\Lambda $ dominates.
\begin{equation}
ds_{-}^2=-(1-\frac 13\Lambda r_{-}^2)dt_{-}^2+\frac 1{(1-\frac 13\Lambda
r_{-}^2)}dr_{-}^2+r_{-}^2(d\theta ^2+sin^2\theta d\phi ^2)
\end{equation}
In the second case when the fermion number density $\rho _0$ is the prevailing
parameter, the interior metric is replaced by the Friedman solution \cite
{ll:pole}. The line element has now the form
\begin{equation}
dl^2=-\frac{dr^2}{(1-{(\frac{r}{a})}^2)}+r^2(d\theta ^2+\sin ^2\theta d\phi ^2)
\end{equation}
where $a$ is a curvature radius. After performing the following
substitution $r=a\sin \chi $ where $\chi $ changes from $0$ to $\pi $, the line
element
takes the form
\begin{equation}
dl^2=a^2\{ d\chi ^2+\sin ^2\chi (d\theta ^2+\sin ^2\theta d\phi ^2)\}
\end{equation}
For space with  constant, positive curvature and for
$a=\alpha r_g$  where $\alpha$ is the dimensionless parameter,
the metric can be written in the following way
\begin{equation}
ds^2=-dt^2+\alpha ^2r_g^2d\chi ^2+\alpha ^2r^2(d\theta ^2+\sin ^2\theta
d\phi ^2).
\end{equation}
Because
\begin{equation}
r_gd\chi =\frac{dr}{1-\frac{r^2}{r_g^2}}
\end{equation}
finaly one can get the metric
\begin{equation}
ds^2=-dt_{-}^2+\alpha ^2 \left( \frac{dr_{-}^2}{(1-\frac{r_{-}^2}{r_g^2})}%
+r_{-}^2(d\theta ^2+\sin ^2\theta d\phi ^2)\right) .
\end{equation}
The time evolution of the scale factor $a$ inside the soliton star resembles
cosmological evolution of the universe but on a different scale. Inside the
soliton star we have a nonvanishing cosmological constant $\Lambda =\kappa
B$ and  fermions which at  rough approximation are treated as dust
density $\rho _o$. The space-time in the soliton star will
evolve according to the Einstein equation which gives, as in cosmology, the
scale evolution equation 
\begin{equation}
\dot{a}^2+1=\frac 43\pi G\rho _0a^2+\frac 13\Lambda a^2
\end{equation}
It is possible to interpret the time
evolution of the metric inside the soliton as a movement of the fictitious
particle with the vanishing energy under the influence of some effective
potential 
\begin{equation}
\frac{1}{2}\dot{a}^2+U_{eff}(a)=E
\end{equation}
For $E=0$ 
\begin{equation}
U_{eff}(a)=\frac 12-\frac{8\pi G}6({\rho }_0a_0^3)\frac 1a-\frac{8\pi G}%
6Ba^2
\end{equation}
where the dust mass $M_d=\rho _0V$ with $V=2\pi ^2a^3$. The ''steady state''
solution is obtained for
\begin{equation}
a_0=(\frac{GM_d}{\pi \Lambda })^{\frac 13}=(\frac{M_d}{8\pi ^2B})^{\frac 13}
\end{equation}
It is convenient to introduce the dimensionless potential 
\begin{equation}
U_{eff}(x)=\frac 12-A(\frac 1x+\frac 12x^2)
\end{equation}
which depends only on one parameter $A$ and $a=a_0x$
\begin{equation}
A=(\frac{GM_d}{3\pi a_0}).
\end{equation}
The critical point $A_c = \frac{1}{3}$ represents the maximum of the potential $U_{eff}(x)$%
Fig.2 presents the form of the potential function $U_{eff}(x)$ for two
values of parameters $A$; $A=A_c$ and $A=\frac 23$. Results obtained for
different values of parameter $B$ are presented in Table1.
\begin{center}
\begin{tabular}{|c|c|}
\hline
$B=(0.1 GeV)^4$ & $B=(1 GeV)^4$ \\ \hline
$a_{0}$ = 1036 km & $a_{0}$ = 0.48 km \\ \hline
A = 0.001524 & A = 0.3284 \\ \hline
\end{tabular}
\end{center}
The metric evolution is determined by the critical value of parameter $A_c$
which produces the critical scale $a_c$. If $A>A_c$, the cosmological constant
$\Lambda $ can be neglected and the evolution is exactly the same as in the
Friedman universe with gravitational collaps as a result. For the soliton
star with $B=56MeV/fm^3$ we obtain $\rho _c\sim 10^{11}g/cm^3$ which is
below the neutron star density. If $A<A_c$, the cosmological constant $%
\Lambda $ dominates and space-time expands. The empty soliton will expand
exponentially according to the de Sitter solution with \textit{the Hubble
constant} 
\begin{equation}
H=\sqrt{\frac{8\pi G}3B}.
\end{equation}
The bag constant B=56 MeV/$fm^3$ determines the value of the parameter $t_{H}$
\begin{equation}
t_H=\frac 1{H}\sim 10^5s\sim 1day.
\end{equation}
In order to derive the equations of motion for the surface of a soliton star
we have to obtain the junction conditions between the inner and outer region
of the soliton. Let $\Sigma $ be a three-dimensional spacetime hypersurface
swept out by the soliton surface. It is convenient to choose the Gaussian
normal coordinates in the neighbourhood of $\Sigma $. Coordinates of any
point $p$ belonging to the neighbourhood $N$ of $\Sigma $ are given by $%
x_\mu =(x_i,\eta )$, $x_i=(\tau ,\theta ,\varphi )$ where $\varphi $ and $%
\theta $ are angular variables, $\tau $ the proper time of an observer
comoving with the domain wall, $\eta $ is the distance from $\Sigma $ to $p$
along the geodesics which are orthogonal to $\Sigma $ \cite{gu:infl}. We also
  define $\xi ^\mu $, as a unit vector field normal to $\eta =$ const
hypersurface. The three dimensional metric intrinsic to the hypersurface is
\begin{equation}
h_{\mu \nu }=g_{\mu \nu }-\xi _\mu \xi _\nu 
\end{equation}
where $g_{\mu \nu }$ is the four metric of the spacetime. According to
Israel \cite{is:zszy}, in this case the 3-tensor $\gamma _{ij}$ defined as
\begin{equation}
\gamma _{ij}=K_{ij}^{+}-K_{ij}^{-} \neq 0
\end{equation}
where $K_{ij}$ is the extrinsic curvature 3-tensor of $\eta =$ const
hypersurface. This tensor can be calculated using the following relations
\begin{equation}
K_{ij}=-\Gamma _{ij}^\alpha =\frac 12\frac{\partial {g_{ij}}}{\partial {%
x^\alpha }}.
\end{equation}
We are interested in the situation when the energy-momentum tensor $T_{\mu
\nu }$ of the four-dimensional spacetime has a $\delta $-function
singularity on the hypersurface and can be expressed in the following form 
\cite{gu:infl} 
\begin{equation}
T_{\mu \nu }=S_{\mu \nu }(x_i)\delta (\eta )
\end{equation}
with $S_{\mu \nu }$ as the surface energy tensor. Taking this into account
the extrinsic curvature $K_{ij}$ possesses a jump discontinuity across the
hypersurface. Adopting the Gauss-Codazzi formalism \cite{gu:infl} the $%
\gamma _{ij}$ tensor takes the form 
\begin{equation}
\gamma _{ij}=-8\pi G(S_{ij}-\frac 12g_{ij}S)
\end{equation}
where $S_{ij}$ is defined as the integral of the energy-momentum tensor
through the thickness of the wall. 
\begin{equation}
S_{ij}=\lim\limits_{\epsilon \rightarrow 0}\int\limits_0^\epsilon T_{ij}dx^1
\end{equation}
We will limit our consideration to the following form of the surface energy
tensor \cite{gu:infl}
\begin{equation}
S_{\mu \nu }=\sigma (\tau )u_\mu u_\nu -\zeta (\tau )(h_{\mu \nu }+u_\mu
u_\nu )
\end{equation}
where $\sigma $ is the surface energy density of the domain wall, $\zeta $
the surface tension and $u_\mu $  the four-velocity of the wall. For
the domain wall $\sigma =\zeta $ and
\begin{equation}
S_{\mu \nu }=-\sigma h_{\mu \nu }  \label{eq:pow}
\end{equation}
In order to describe the dynamics of the wall which separates the interior
and the exterior of the soliton star we have to calculate the appropriate
components of the extrinsic curvature; $K_{ij}^{+}$ - the extrinsic
curvature for the Schwarzschild metric and $K_{ij}^{-}$ for the metric which
describes the space inside the soliton. Taking into account the form of $%
S_{\mu \nu }$ (\ref{eq:pow})
\begin{equation}
\gamma _j^i=-4\pi G\sigma \delta _j^i
\end{equation}
For the system described by the Lagrange function (\ref{eq:lag}) it is
possible to interpret the Einstein equation having the following form inside
the soliton\\
\begin{equation}
R_{\mu \nu }-\frac 12g_{\mu \nu }R=\kappa Bg_{\mu \nu }+\kappa T_{\mu \nu
}(\psi )
\end{equation}
or\\ 
\begin{equation}
R_{\mu \nu }-\frac 12g_{\mu \nu }R-\Lambda g_{\mu \nu }=\kappa T_{\mu \nu
}(\psi )
\end{equation}
with $\Lambda =\kappa B$ as the Einstein equation with the cosmological
constant different from zero and with the fermion energy-momentum tensor. In
the case when $T_{\mu \nu }(\psi )=0$ (empty bag) the Einstein equation is
reduced to the form with the solution expanding according to the de Sitter
law. It is interesting to present the dynamical equation of the soliton
surface. Owing to the
spherical symmetry of $\Sigma $ the metric on the domain wall can be
presented in the following way 
\begin{equation}
ds^2=-d\tau ^2+R(\tau )^2(d\theta ^2+\sin ^2\theta d\phi ^2)
\end{equation}
where $\tau $ is a proper time along the world-line ($\theta ,\phi =const$)
of a dust particle. We adopt this metric as a matching condition of the
interior and the exterior metric on the soliton surface. Now using the
method presented in Israel's paper \cite{is:zszy} we obtain the dynamical
equation for the soliton surface
\begin{equation}
(1-\frac 13\Lambda R^2+{\dot{R}}^2)^{\frac 12}-(1-\frac{r_g}R+{\dot{R}}%
^2)=4\pi \sigma R
\end{equation}
In all these equations $\dot{R}=\frac{dR}{d\tau }$. This equation agrees with
that obtained by A.Aurilia \cite{au:kola}. Once more the surface dynamics
can be illustrated as fictitious particle motion with the total energy equal
to zero in the effective potential field. 
\begin{equation}
\frac 12\dot{R}^2=-V_{eff}(R)
\end{equation}
where $V_{eff}(R)$ can be expressed as
\begin{eqnarray}
V_{eff}(R) &=&\frac 12 \{ 1-R^2(\frac{B}{3\sigma }+2\pi G\sigma )^2  \nonumber \\
&-&\frac MR(\frac B{6\pi G\sigma ^2}+1)-\frac{M^2}{16\pi ^2\sigma ^2R^4} \}
\end{eqnarray}
substituting $R=xr_g$, $r_g$ being the gravitational radius, the
effective potential $V_{eff}(R)$ takes the following form
\begin{eqnarray}
V_{eff} &=&\frac 12 \{ (1-x^2r_g^2(\frac B{3\sigma }+2\pi G\sigma )^2  \nonumber
\\
&-&\frac 1{2x}(\frac B{6\pi G\sigma ^2}+1)-\frac 1{(8\pi Gr_g\sigma
)^2x^4}\}
\end{eqnarray}
We can also present the mass $M$ of the soliton 
\begin{equation}
M=\frac 43\pi R^3B+4\pi R^2\sigma (1-\frac 83\pi GBR^2+\dot{R}^2)^{\frac
12}-8{\pi }^2G{\sigma }^2R^3
\end{equation}
The subsequent terms in the expression for the mass are the volume energy
and the surface energy with the relativistic correction and the surface-surface binding
energy.
Fig.3 presents the form of the potential $V_{eff}(x)$
which depends only on one variable $x$. The soliton described in this paper
is not stable and such a configuration tends to obtain the equlibrium state by
the emission of particles. Phenomenological description of this situation
can be made by introducing the term identified as  friction.\\ In the
vicinity of point $x_c$ which corresponds to the maximum of the potential $%
V_{eff}(x)$, the function $V_{eff}(x)$ can be approximated in the following
way.
\begin{equation}
V_{eff,a}(x)=-\frac 12u_0+\frac 12\omega ^2(x-x_0)^2
\end{equation}
On Fig.3 the function $V_{eff,a}(x)$ is represented by the broken line.
Substituting $y=x-x_0$ one can write
\begin{equation}
\dot{y}^2=\omega ^2(y+a^2)
\end{equation}
where $a^2=\frac{u_0}{\omega ^2}$. The general solutions of this equation have
the following form
\begin{equation}
y_{1}=\frac 12\{e^{-t_0}e^{\omega t }(-a^2+e^{2t_0}e^{-2\omega t })\}
\end{equation}
\begin{equation}
y_{2}=\frac 12\{e^{-t_0}e^{-\omega t }(-a^2+e^{2t_0}e^{2\omega t })\}
\end{equation}
where $y_{1}$ represents collaps and $y_{2}$ expansion of the soliton.
For $t=0$ 
\begin{equation}
y_0=\frac 12\{e^{-t_0}(-a^2+e^{2t_0})\}=\frac 12\{-a^2e^{-t_0}+e^{t_0}\}
\end{equation}
so for this moment 
\begin{equation}
x_p=x_0+\frac 12\{-a^2e^{-t_0}+e^{t_0}\}
\end{equation}
Substituting $e^{t_0}=az$ 
\begin{equation}
x_p=x_0+\frac 12\frac az(z^2-1)
\end{equation}
The mass $M$ of the soliton is determined by the initial value $x_p$. The
dimensionless parameter $m=\frac M{M_{\odot }}$ 
\begin{equation}
m=\frac 43\pi bx_p^3+4\pi sx_p^2
\end{equation}
The equation of motion takes the following form
\begin{equation}
\ddot{x}+b\dot{x}=-\frac{\partial V_{eff,a}}{\partial x}=\omega ^2(x-x_c) \label{eq:mot}
\end{equation}
We can analyze the solution of the equation (\ref{eq:mot}) in momentum
space Fig.4. Solutions where $p>0, p=\dot{x}$, describe the trajectories representing
the expantion of the whole system
whereas for solutions with $%
p<0$ the trajectories lead to the collaps.
Thus the values of the initial conditions are responsible either for the
expansion or the collapse of the soliton.
There is only one critical trajectory for which the whole system, after an
infinite time, tends to the stable configuration $M=0.92M_{\odot }$ and $%
R=16.33$ km.
Different types of trajectories are presented
on Fig.4. The vertical stright line marks the position of the Schwarzschild horizon,
two hyperbolas correspond to the trajectories if there is no friction in the
equation of motion, broken line presents the
trajectory which passes very close to the critical one. The stright line
shows the position of the critical trajectory.
The analysis of this figure leads to the conclusion that the system reaches stable configuration
before reaching
the Schwarzschild horizon.
Such a configuration corresponds to the situation when the
soliton obtaines the state of equilibrium between the emission of particles
from the surface and the de Sitter expansion.
Fig.5 shows the numerical solution of the equation of motion for the
potential function $V_{eff}(x)$ and this solution confirms the
analysis which was carried out on the basis of Fig.4.
\\ Let us now concentrate on
the situation when the cosmological constant $\Lambda $
is neglected inside the soliton. It has been already mentioned that this very solution
corresponds with the case when the bag constant $B=0$ and represents the
degenerate vacuum state. The confinement is only due to the surface tension $\sigma$.
This situation appears in the case of the Lee-Pang star \cite{lp:star}.
The necessary and sufficient conditions for reality of this system
is both the existence of the nontopological soliton which is guaranteed
by the scalar field $\Phi$, the fermion field and the gravitational
field. The soliton star in this case consists of a large interior
and a much thinner surface shell of the width $\sim \mu^(-1)$, where
$\mu$ is the mass of the scalar field $\Phi$.
For the value of surface tension $\sigma =\frac{1}{6}(30 GeV)^3$ the
following parameters of the soliton star
was obtained: mass$ M \sim 10^13 M_{\odot}$ and radius $R \sim 1$ light year.\\
The exterior geometry is still described by the
Schwarzschild line element whereas the interior geometry is described by the
Robertson-Walker metric. If fermions inside the soliton are treated in the
rough approximation as dust with the density $\rho _0$, the energy momentum
tensor $T_{\mu \nu }$ takes the form 
\begin{equation}
T_{\mu \nu }=\rho _0u_\mu u_\nu 
\end{equation}
where $u_\mu $ is the four-velocity vector of the particle. Once more the
Israel's equation for the motion of the soliton surface separating the two
metrics was obtained 
\begin{equation}
h^{-1}(1+h{\dot{R}}^2)^{\frac 12}-(1-\frac{2GM}R+{\dot{R}}^2)^{\frac
12}=4\pi \sigma GR
\end{equation}
where $h=\frac{\alpha^2}{1-\frac{R}{R_g}}$, $\alpha$ is the scale factor.
In order to study the dynamics of the soliton surface we identified the
motion of the domain wall with the motion of a unit mass particle with
vanishing total energy moving under the influence of the potential. The form
of the potential function $W(\Phi)$ is presented on Fig.6. Closer analysis of
this function reveals that once again if we introduce the friction
the whole system behaves just like in the de Sitter case. One can notice the
similarity with the results obtained in the previous paragraph concerning
the analysis of the function $V_{eff}(x)$. However, this time the critical
trajectory reaches the Schwarzschild horizon  $R=r_g$.
Thus in this case the soliton collapses to the black hole.

\section{Conclusions}
The analysis of the soliton surface dynamics has been performed in this paper.
It may be concluded that the star either collapses into a black
the whole system is characterized by the following parameters the bag constant $B$,
the surface tension $\sigma$,and the fermion number density $\rho_{0}$.
If the cosmological constant $\Lambda$ is the prevailing parameter
the space time inside the soliton can be described by the de Sitter solution.
When  $\rho_{0}$ prevails the Friedman solution is obtained.
It turned out in the de Sitter  case there exists the stable configuration
described by the macroscopic parameters: the mass $M=0.92 M_{\odot}$ and the radius $R=16.33 $ km.
The radius of this configuration is bigger than the gravitational one and chartacteristic
of the whole system resambles the ones that can be observed in neutron stars.
In the case of Friedman solution the soliton which is characterized by the mass $M\sim 10^{13}M_{\odot }$
collapses exactly at the Schwarzschild horizon,
so in the result we obtain the black hole.
\section*{Acknowledgements}
This project was sponsored by the KBN Grant 2 P304 022 06.\\
All calculations and figures were done using {\it Mathematica}
(L2327-0416, {\it Mathematica}, Wolfram Research, Inc. Champaign,USA).

\newpage

\begin{center}
\textbf{Figure captions}
\end{center}
{\bf Fig.1.} The form of the potential function $U(\Phi)$.\\
{\bf Fig.2.} The effective potential $U_{eff}(x)$ for two values of parameter $A$:
the continues line corresponds to the $A=A_c$ wherease dotted line to $A=\frac{2}{3}$.\\
{\bf Fig.3.} The form of the effective potential $V_{eff}(x)$ - continues line,
dotted line indicates the possition of the function $V_{eff,a}(x)$.\\
{\bf Fig.4.} The trajectories for different types of solutions presented
in the phase space.\\
{\bf Fig.5.} Numerical solution of the equation of motion \ref{eq:mot}\\
{\bf Fig.6.} The effective potential $W(\Phi)$.\\
\newpage
\epsfbox{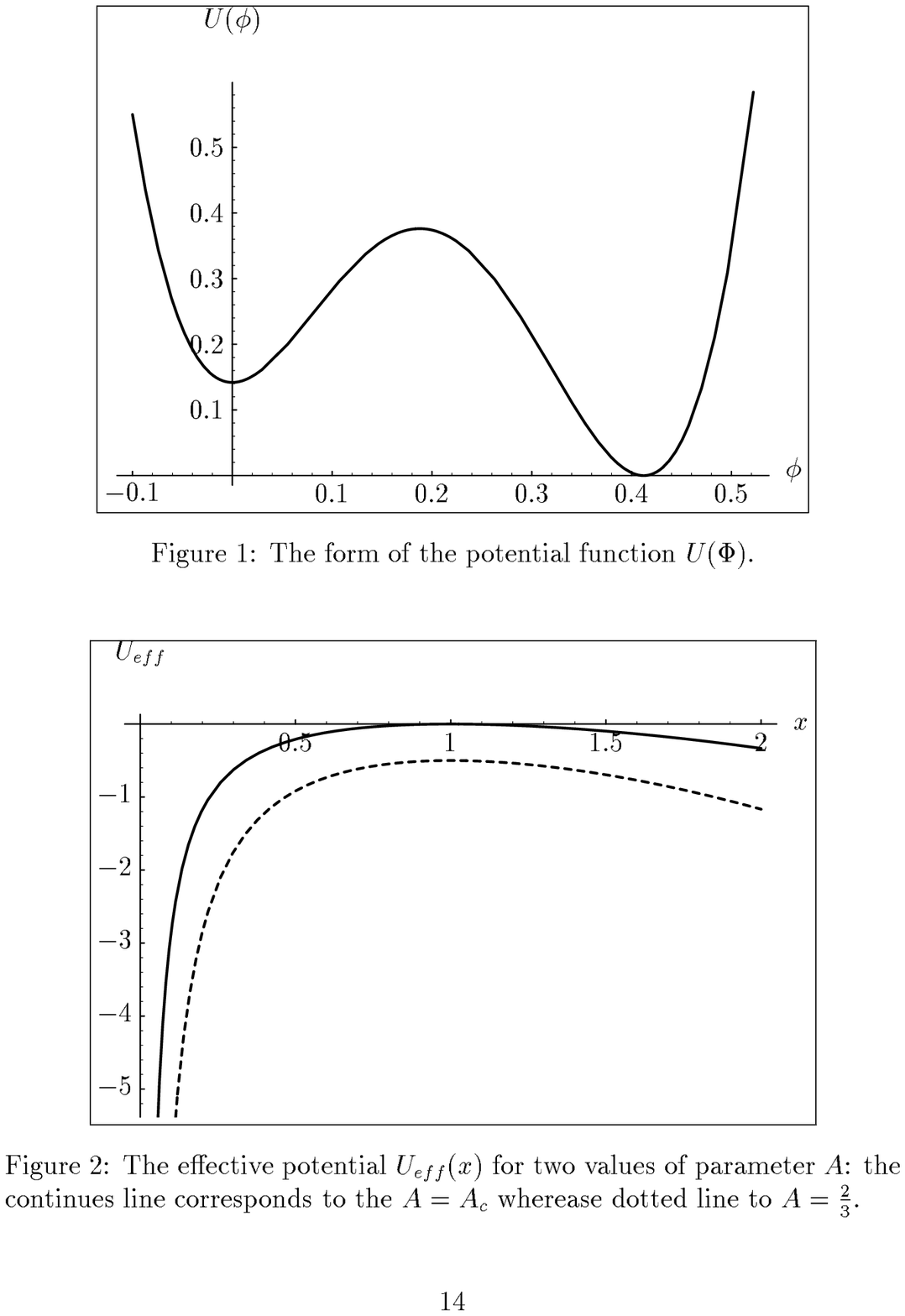}
\newpage
\epsfbox{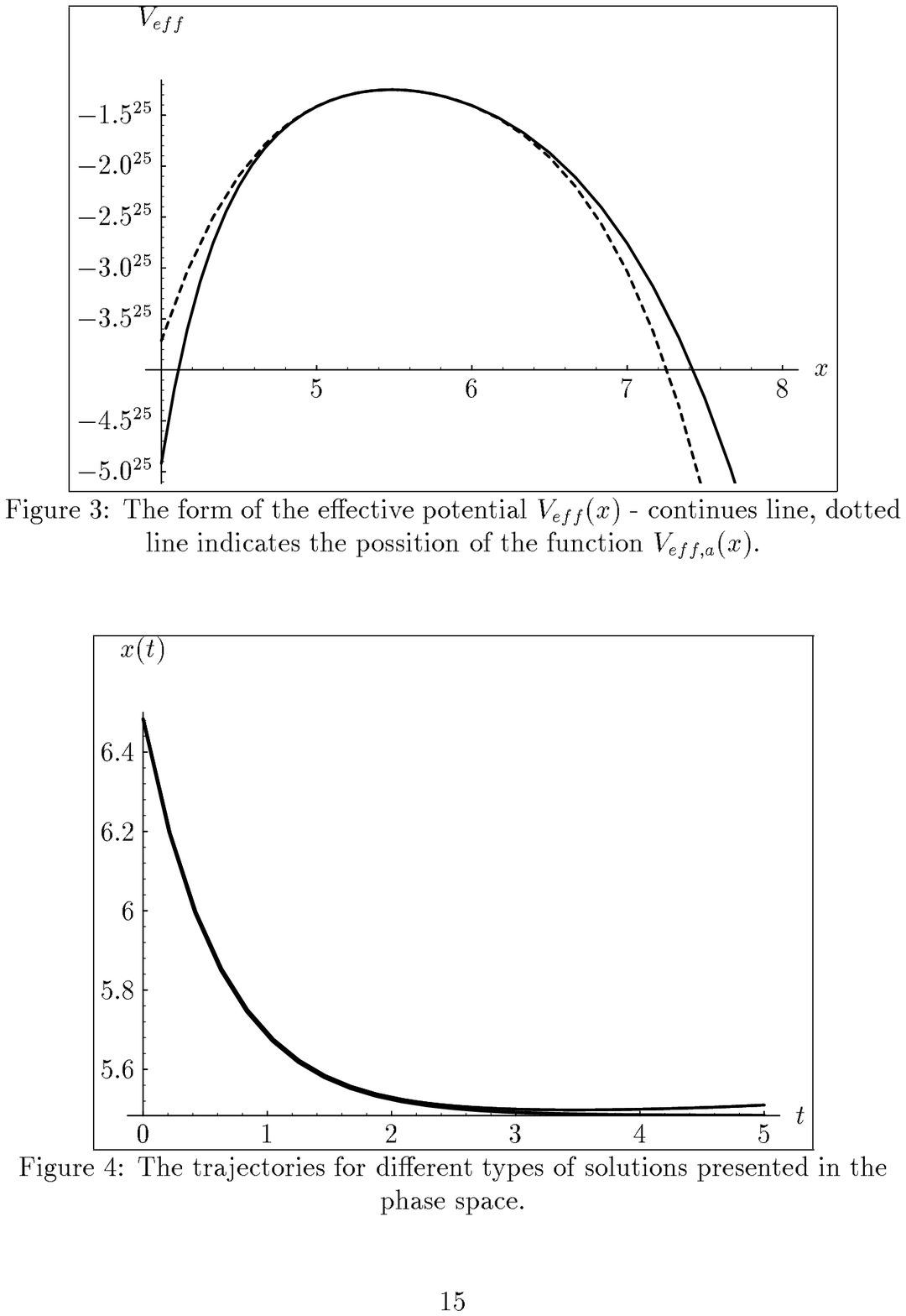}
\newpage
\epsfbox{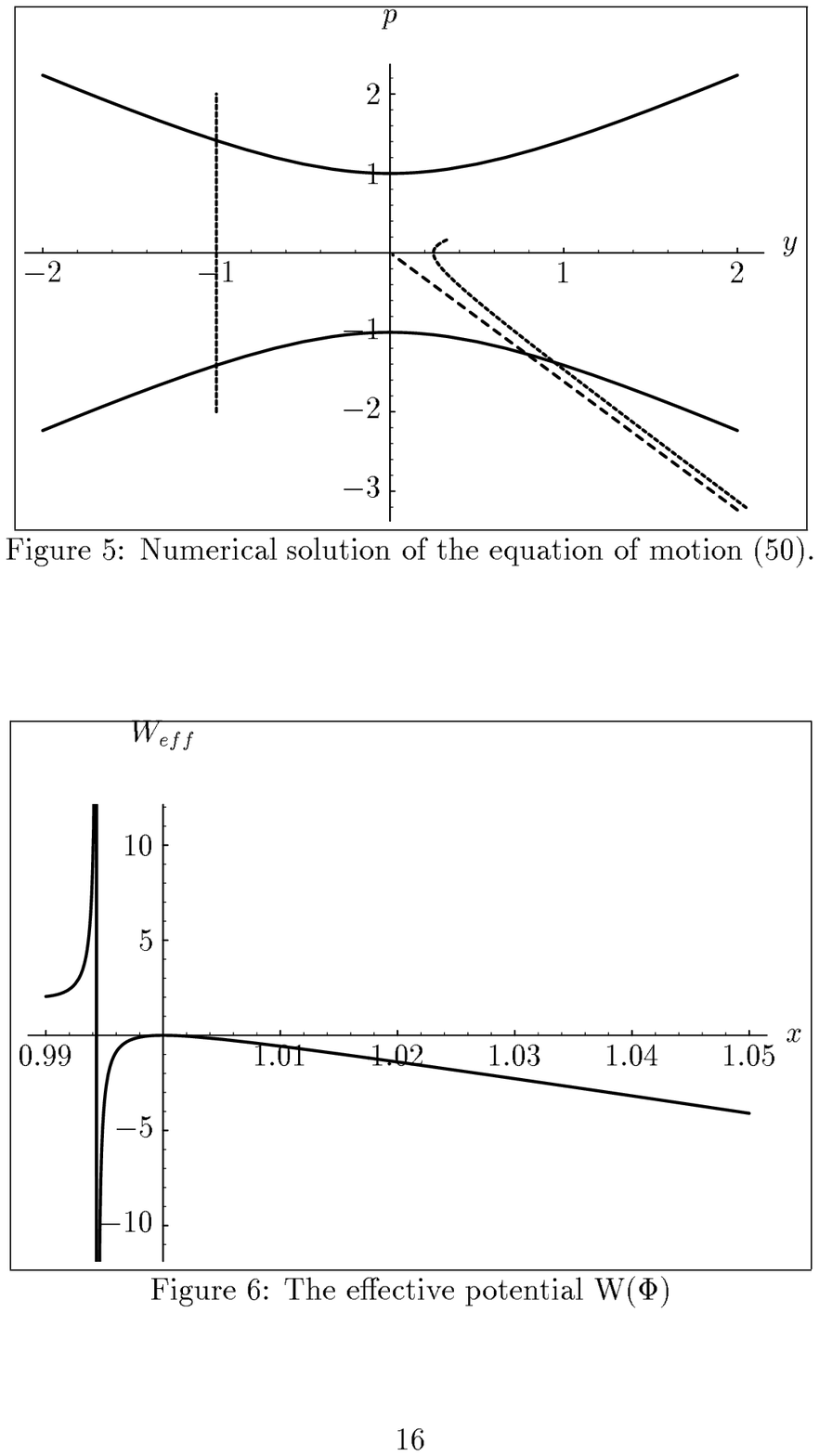}
\end{document}